\documentclass[twocolumn,showpacs,preprintnumbers,amsmath,amssymb,prb]{revtex4}

\usepackage{graphicx}
\usepackage{dcolumn}
\usepackage{longtable}
\usepackage{bm}

\begin{document}

\title{Structural relaxation around substitutional Cr$^{3+}$ in MgAl$_{2}$O$_{4}$} 

\author{Am\'{e}lie Juhin} 
\email{amelie.juhin@impmc.jussieu.fr} 
\affiliation{Institut de Min\'eralogie et 
 de Physique des Milieux Condens\'es, UMR CNRS 7590\\
 Universit\'e Pierre et Marie Curie, Paris 6\\
 140 rue de Lourmel, F-75015 Paris, France}
\author{Georges Calas}
\affiliation{Institut de Min\'eralogie et 
 de Physique des Milieux Condens\'es, UMR CNRS 7590\\
 Universit\'e Pierre et Marie Curie, Paris 6\\
 140 rue de Lourmel, F-75015 Paris, France}
\author{Delphine Cabaret} 
\affiliation{Institut de Min\'eralogie et 
 de Physique des Milieux Condens\'es, UMR CNRS 7590\\
 Universit\'e Pierre et Marie Curie, Paris 6\\
 140 rue de Lourmel, F-75015 Paris, France}
\author{Laurence Galoisy}
\affiliation{Institut de Min\'eralogie et 
 de Physique des Milieux Condens\'es, UMR CNRS 7590\\
 Universit\'e Pierre et Marie Curie, Paris 6\\
 140 rue de Lourmel, F-75015 Paris, France}
\author{Jean-Louis Hazemann}
\email{hazemann@grenoble.cnrs.fr} 
\affiliation{Laboratoire de Cristallographie, CNRS, 25 avenue des Martyrs, BP 166, 38042 Grenoble cedex 9, France}

\date{\today}

\begin{abstract}
The structural environment of substitutional Cr$^{3+}$  ion in MgAl$_{2}$O$_{4}$ spinel has been investigated by  Cr K-edge Extended X-ray Absorption Fine Structure (EXAFS) and X-ray Absorption Near Edge Structure (XANES) spectroscopies. First-principles computations of the structural relaxation and of the XANES spectrum have been performed, with a good agreement to the experiment. The Cr-O distance is close to that in MgCr$_{2}$O$_{4}$, indicating a full relaxation of the first neighbors, and the second shell of Al atoms relaxes partially. These observations demonstrate that Vegard's law is not obeyed in the MgAl$_{2}$O$_{4}$-MgCr$_{2}$O$_{4}$ solid solution. Despite some angular site distortion, the local D$_{3\mathrm d}$  symmetry of the B-site of the spinel structure is retained during the substitution of Cr for Al. Here, we show that the relaxation is accomodated by strain-induced bond buckling, with angular tilts of the Mg-centred tetrahedra around the Cr-centred octahedron. By contrast, there is no significant alteration of the angles between the edge-sharing octahedra, which build chains aligned along the three four-fold axes of the cubic structure.
\end{abstract} 


\maketitle

\section{Introduction}

Most multicomponent materials belong to complete or partial solid solutions. The presence of chemical substitutions gives rise to important modifications of the physical and chemical properties of the pure phases. For instance, the addition of a minor component can improve significantly the electric, magnetic or mechanical behaviour of a material.\cite{Levy,Lau,Frenk} Another evidence for the presence of impurities in crystals comes from the modification of optical properties such as coloration. Transition metal ions like Cr$^{3+}$ cause the coloration of wide band gap solids, because of the splitting of 3$d$-levels under the action of crystal field.\cite{Burns} Despite the ubiquitous presence of substitutional elements in solids, their accommodation processes and their structural environment are still discussed,\cite{Blun} since they have important implications. For example, the interpretation of the color differences between Cr-containing minerals (e.g. ruby, emerald, red spinel) requires to know the structural environment of the coloring impurity. \cite{Burns,Garcia,Moreno,Gaud1} The ionic radius of a substitutional impurity being usually different from that of the substituted ion, the accommodation of the mismatch imposes a structural relaxation of the crystal structure. Vegard's law states that there is a linear relationship between the concentration of a substitutional impurity and the lattice parameters, provided that the substituted cation and impurity have similar bonding properties. Chemically selective spectroscopies, like Extended X-ray Absorption Fine Structure (EXAFS), have provided evidence that diffraction studies of solid solutions give only an average vision of the microscopic states and that Vegard's law is limited.\cite{Lan,Mart,Gal} Indeed, a major result concerns the existence of a structural relaxation of the host lattice around the substitutional cation. This implies the absence of modification of the site occupied by a doping cation, when decreasing its amount in a solid solution. 
This important result has been observed in various materials, including III-V semi-conductors or mixed salts: \cite{Mikk,Frenk2} e. g., in mixed alkali halides, some important angular buckling deviations have been observed.\cite{Frenk2} Recently, the use of computational tools, as a complement of EXAFS experiments, has been revealed successful for the study of oxide/metal epilayers.\cite{Lamb} In oxides containing dilute impurities, this combined approach is mandatory. It has been recently applied to the investigation of the relaxation process around Cr dopant in corundum: in the $\alpha$-Al$_{2}$O$_{3}$ - $\alpha$-Cr$_{2}$O$_{3}$ system, the radial relaxation was found to be limited to the first neighbors around Cr, while the angular relaxation is weak.\cite{Gaud1,Gaud2} 

In this work, we investigate the relaxation caused by the substitution of Al$^{3+}$ by Cr$^{3+}$ in spinel MgAl$_{2}$O$_{4}$, which gives rise to a solid solution, as observed for corundum $\alpha$-Al$_{2}$O$_{3}$. The spinel MgAl$_{2}$O$_{4}$ belongs to an important range of ceramic compounds, which has attracted considerable interest among researchers for a variety of applications, great electrical, mechanical, magnetic and optical properties.\cite{Thib}  The spinel structure is based on a cfc close-packing, with a Fd$\bar{3}$m space group symmetry. Its chemical composition is expressed as AB$_2$X$_4$, where A and B are tetrahedral and octahedral cations, respectively, and X is an anion. These two types of cations define two different cationic sublattices, which may  induce a very different relaxation process than in corundum. In the normal spinel structure, the octahedra host trivalent cations and exhibit D$_{3\mathrm d}$ site symmetry. This corresponds to a small distortion along the [111] direction, arising from a departure of the position of oxygen ligands from a cubic arrangement. Small amounts of chromium oxide improve the thermal and mechanical properties of spinel.\cite{Levy} A color change from red to green is also observed with increasing Cr-content. In this article, we report new results on the local geometry around Cr$^{3+}$ in spinel MgAl$_{2}$O$_{4}$, using a combination of EXAFS and X-ray Absorption Near Edge Structure (XANES). The experimental data are compared to those obtained by theoretical calculations, based on the Density Functional Theory in the Local Spin Density Approximation (DFT-LSDA): this has enabled us to confirm the local structure around substitutional Cr$^{3+}$ and investigate in detail the radial and angular aspects of the relaxation.

The paper is organized as follows. Section II is dedicated to the methods, including the sample description (Sec. II A), the X-ray absorption measurements and analysis (Sec. II B), and the computational details (Sec. II C). Section III is devoted to the results and discussion. Conclusions are given in Sec. IV.

\section{Materials and methods}
\subsection{Sample description}
Two natural gem-quality red spinel single crystals from Mogok, Burma (Cr-1, Cr-2) were investigated. They contain respectively 70.0, 71.4 wt \%-Al$_{2}$O$_{3}$, 0.70, 1.03 wt\%-Cr$_{2}$O$_{3}$ and 26.4, 25.3 wt\%-MgO. These compositions were analyzed using the Cameca SX50 electron microprobe at the CAMPARIS analytical facility of the Universities of Paris 6/7, France. A 15 kV voltage with a 40 nA beam current was used. X-ray intensities were corrected for dead-time, background, and matrix effects using the Cameca ZAF routine. The standards used were  $\alpha$-Al$_{2}$O$_{3}$,  $\alpha$-Cr$_{2}$O$_{3}$ and MgO. 
\subsection{X-ray Absorption Spectroscopy measurements and analysis}
Cr K-edge (5989 eV) X-ray Absorption Spectroscopy (XAS) spectra were collected at room temperature at beamline BM30b (FAME), at the European Synchrotron Radiation Facility (Grenoble, France) operated at 6 GeV. The data were recorded using the fluorescence mode with a Si (111) double crystal and a Canberra 30-element Ge detector.\cite{Proux} We used a spacing of 0.1 eV  and of 0.05 \AA$^{-1}$, respectively in the XANES  and EXAFS regions. Data treatment was performed using ATHENA following the usual procedure and the EXAFS data were analyzed using IFEFFIT, with the support of ARTEMIS. \cite{New} The details of the fitting procedure can be found elsewhere.\cite{Rav} An uvarovite garnet, Ca$_{3}$Cr$_{2}$Si$_{3}$O$_{12}$, was used as model compound to derive the value of the amplitude reduction factor S$_{0}^{2}$ (0.81) needed for fitting. For each sample, a multiple-shell fit was performed in the q-space, including the first four single scattering paths: the photoelectron is backscattered either by the first (O), second (Al or Cr), third (O) or fourth (Mg) neighbors. Treating identically the third and fourth paths, we used a unique energy shift $\Delta$e$_{0}$ for all paths, three different path lengths R and three independent values of the Debye-Waller factor $\sigma^{2}$. In a first step, the number of neighbors N was fixed to the path degeneracy. 
In a second time, a single amplitude parameter was fitted for the last three shells, assuming a proportional variation of the number of atoms on each shell. 
\subsection{Computations}
\subsubsection{Structural relaxation}
In order to complement the structural information from EXAFS, a simulation of the structural relaxation was performed to quantify the geometric surrounding around an isolated Cr$^{3+}$. The calculations were done in a neutral supercell of MgAl$_{2}$O$_{4}$, using a first-principles total energy code based on DFT-LSDA.\cite{Paratec} We used Plane Wave basis set and norm conserving pseudopotentials\cite{Trou} in the Kleiman Bylander form.\cite{Klein} 
For Mg, we considered 3$s$, 3$p$, 3$d$ as valence states (core radii of 1.05 a.u, $\ell$=2 taken as local part) and those of Ref.\onlinecite{Gaud2} for Al, Cr, O. We first determined the structure of bulk MgAl$_{2}$O$_{4}$. We used a unit cell, which was relaxed with 2$\times$2$\times$2 $k$-point grid for electronic integration in the Brillouin Zone and cut-off energy of 90 Ry. We obtained a lattice constant of 7.953~\AA~ and an internal parameter of 0.263  (respectively -1.6~\%~ and +0.3~\%~ relative to experiment),\cite{Yama} which are consistent with previous calculations.\cite{Thib}. 
In order to simulate the Cr defect, we used a 2$\times$2$\times$2 supercell, built using the relaxed positions of the pure phase. It contains 1 neutral Cr, 31 Al, 16 Mg and 64 O atoms. It was chosen large enough to minimize the interaction between two paramagnetic ions, with a minimal Cr-Cr distance of 11.43~\AA. While the size of the supercell is kept fixed, all atomic positions are relaxed in order to investigate long-range relaxation. We used the same cut-off energy and a single $k$-point sampling. The convergence of the calculation was verified by comparing it to a computation with a 2$\times$2$\times$2 $k$-point grid, and discrepancies in the atomic forces are lower than 0.3 mRy/a.u. In order to compare directly the theoretical bond distances to those obtained by EXAFS spectroscopy, the inital slight underestimation of the lattice constant (systematic within the LDA) \cite{Louie} was removed by rescaling the lattice parameter by -1.6~\%. This rescaling is homothetic and does not affect the relative atomic positions.
\subsubsection{XANES simulations}
As the analysis of the experimental XANES data is not straightforward, ab initio XANES simulations are required to relate the experimental spectral features to the local structure around the absorbing atom.
The method used for XANES calculations are described in Ref. \onlinecite{Taille,Caba}. The all-electron wave-functions are reconstructed within the projector augmented wave framework.\cite{Blochl} In order to allow the treatment of large systems, the scheme uses a recursion method to construct a Lanzcos basis and then compute the cross section as a continued fraction.\cite{Hay,Hay2}
The XANES spectrum is calculated in the electric dipole approximation, using the same first-principles total energy code as the one used for the structural relaxation. It was carried out in the relaxed 2$\times$2$\times$2 supercell (i.e 112 atoms), which contains one Cr atom and results from ab initio energy minimization mentioned in the previous subsection. The pseudopotentials used are the same as those used for structural relaxation, except for Cr. Indeed, in order to take into account the core-hole effects, the Cr pseudopotential is generated with only one 1$s$ electron. Convergence of the XANES calculation is reached for the following parameters: a 70 Ry energy cut-off for the plane-wave expansion, one $k$-point for the self-consistent spin-polarized charge density calculation, and a Monkhorst-Pack grid of 3$\times$3$\times$3 $k$-points in the Brillouin Zone for the absorption cross-section calculation. The continued fraction is computed with a constant broadening $\gamma$=1.1 eV, which takes into account the core-hole lifetime.\cite{Krause} 
\begin{table}[!b]
\caption{Structural parameters obtained from the EXAFS analysis in the R range [1.0-3.1 \AA] for Cr-1 and Cr-2 samples. The energy shifts $\Delta$e$_{0}$ were found equal to 1.3 $\pm$ 1.5 eV. The obtained RF factors were 0.0049 and 0.0045.}
\label{tab:fit}
\begin{ruledtabular}
\begin{tabular}{ccccc}
      & R(\AA) &  N  &  $\sigma^{2}$ (\AA${^2}$) &  \\ \hline
Cr-O  & 1.98 & 6.0 & 0.0031 & Cr-1 \\
  & 1.98 & 6.0 & 0.0026 & Cr-2 \\
Cr-Al  & 2.91 & 5.3 & 0.0032 & Cr-1 \\
  &  2.91& 5.4 & 0.0033 & Cr-2 \\
Cr-O  & 3.39 & 1.8 & 0.0079 & Cr-1 \\
  & 3.37 & 1.8 & 0.0077 & Cr-2 \\
Cr-Mg  & 3.39 & 5.3 & 0.0079 & Cr-1 \\
  & 3.39& 5.4 & 0.0077 & Cr-2 \\
\end{tabular}
\end{ruledtabular}
\end{table}

\begin{figure}[!t]
\includegraphics[width=7.9cm]{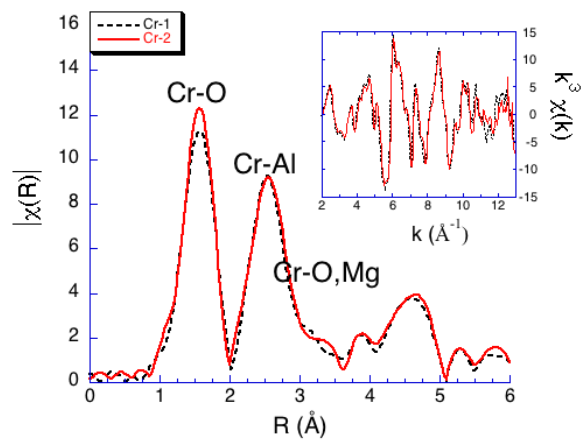}
\caption{\label{fig:1} Fourier-transform of k$^3$-weighted EXAFS function for Cr-1 and Cr-2 samples (dashed and solid lines respectively). Inset: background-subtracted data}
\label{fig:FT}
\end{figure}
\begin{figure}[!b]
\includegraphics[width=7.2cm]{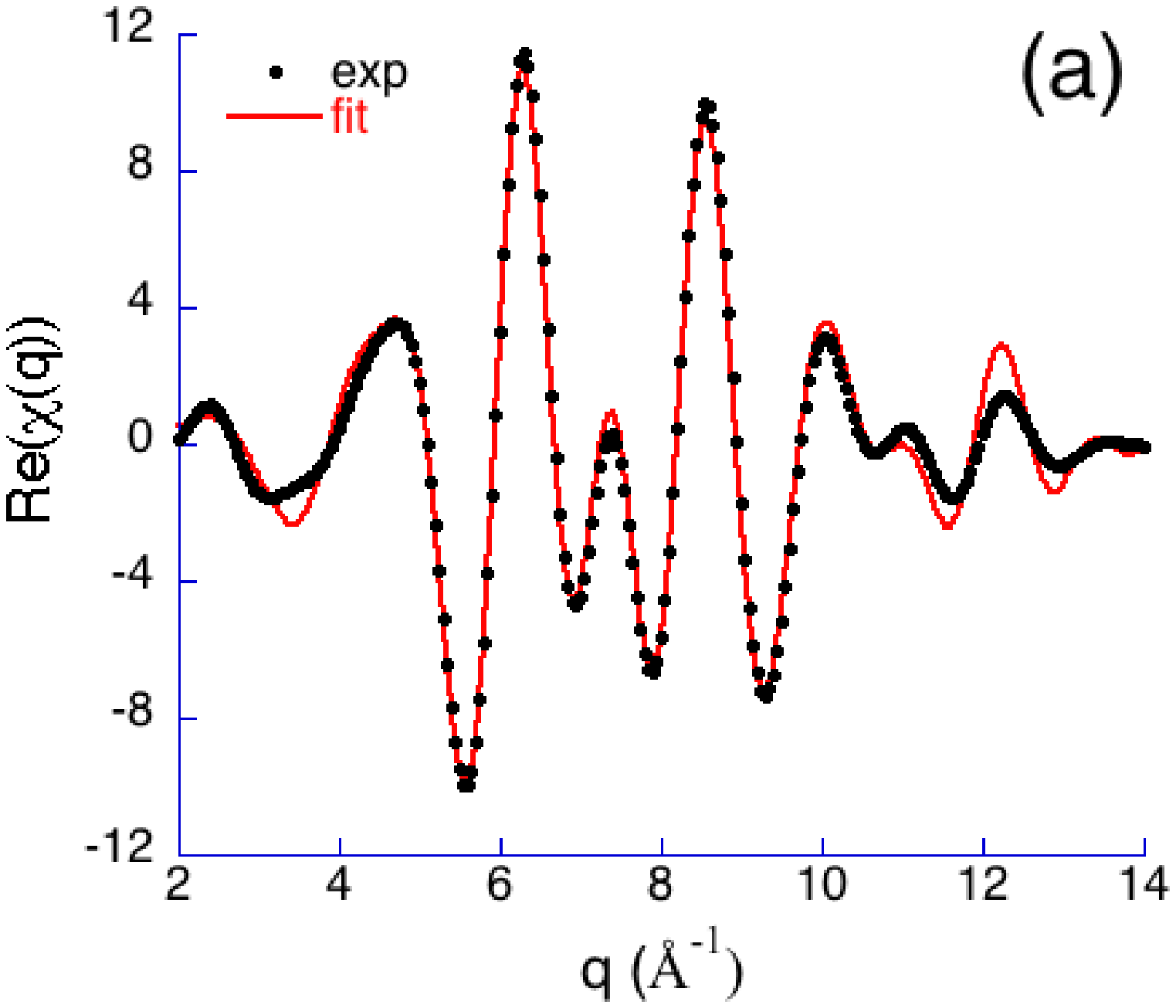}
\includegraphics[width=7.2cm]{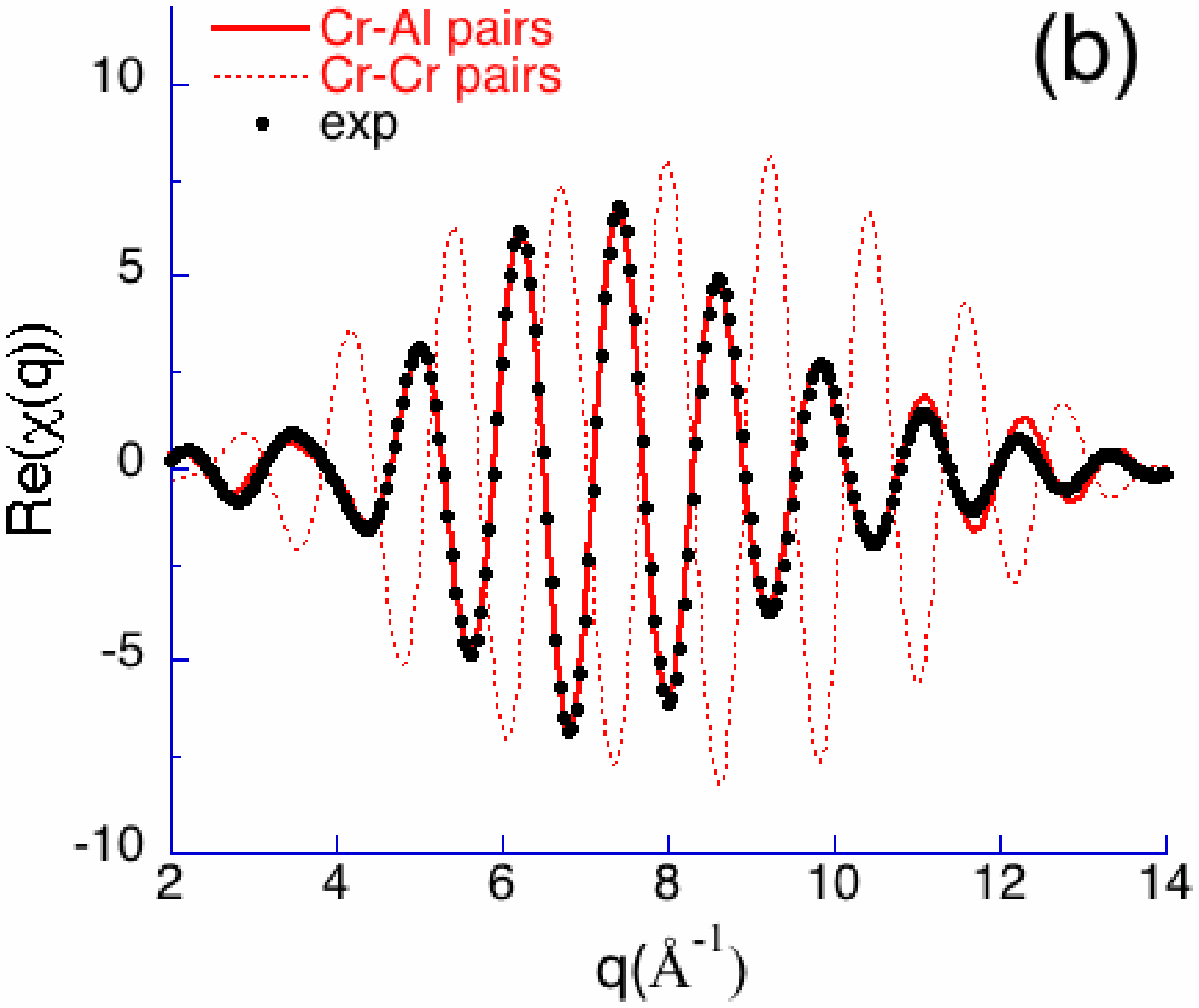}
\caption{\label{fig:2} (a) Inverse-FT of EXAFS data (dots) and fitted signal (solide line) for R=1.0-3.1~\AA. (b) Inverse-FT of EXAFS data (dots) for R=2.0-3.1~\AA, multi-shell fit with Cr-Al pairs (solid line) and theoretical function with Cr-Cr pairs (dashed line) in the same structural model.}
\label{fig:pairs}
\end{figure}

\section{Results and discussion}

\begin{table*}[!t]
\caption{First, second and third neighbor mean distances (in \AA) from central M$^{3+}$ in the different structures considered in this work.}
\label{tab:distances}
\begin{ruledtabular}
\begin{tabular}{ccccc}
      & MgAl$_{2}$O$_{4}$: Cr$^{3+}$ exp    &  MgAl$_{2}$O$_{4}$: Cr$^{3+}$ calc   & MgAl$_{2}$O$_{4}$ exp 
\footnote{from Ref.\onlinecite{Yama}} & MgCr$_{2}$O$_{4}$  exp \footnote{from Ref.\onlinecite{Hill}} \\ \hline
Cr-O  & 1.98 & 1.99 & --- & 1.99 \\
Al-O  & --- & --- & 1.93 & --- \\ \hline
Cr-Al  & 2.91 & 2.88 & --- & --- \\
Cr-Cr  &  --- & --- & --- & 2.95 \\
Al-Al  &  ---& --- & 2.86 & --- \\ \hline
Cr-O  & 3.37 & 3.34 & --- & 3.45 \\
Al-O  & --- & --- & 3.34 & --- \\ \hline
Cr-Mg  & 3.39 & 3.36 & --- & 3.45 \\
Al-Mg  & --- & --- & 3.35 & --- \\ 
\end{tabular}
\end{ruledtabular}
\end{table*}
Figure \ref{fig:FT} shows the k$^3$-weighted experimental EXAFS signals for Cr-1 and Cr-2 samples and the Fourier Transforms (FT) for the k-range 3.7-11.9 \AA$^{-1}$. The similarities observed suggest a close environment for Cr in the two samples (0.70 and 1.03 wt\%-Cr$_{2}$O$_{3}$), which is confirmed by fitting the FT in the R-range 1.0-3.1~\AA~(see Table~\ref{tab:fit}). The averaged Cr-O distance derived from EXAFS data is equal to 1.98~\AA~($\pm$ 0.01~\AA), with six oxygen first neighbors. The second shell is composed of six Al atoms, located at 2.91~\AA~($\pm$ 0.01~\AA). Two oxygen and six magnesium atoms compose the further shells, at distances of 3.38~\AA~ and 3.39~\AA~ ($\pm$ 0.03~\AA). We investigated in detail the chemical nature of these second neighbors, by fitting the second peak on the FT (2.0-3.1~\AA) with either a Cr or an Al contribution, this latter corresponding to a statistical Cr-distribution (Cr/Al $\sim$~0.01). The only satisfactory fits were obtained in the latter case (Fig.~\ref{fig:pairs}).    

\begin{figure}[!b]
\includegraphics*[width=6cm]{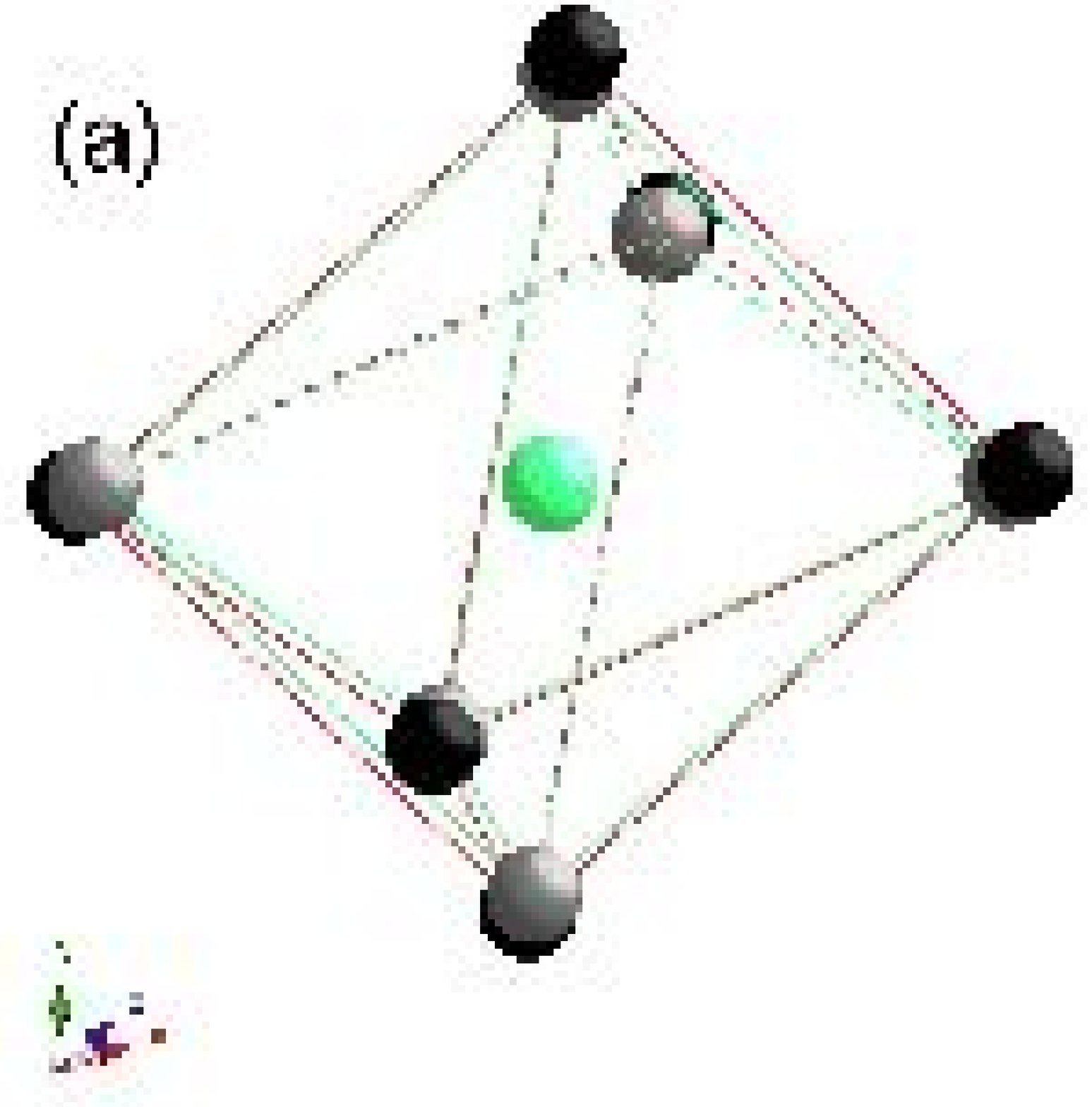}
\includegraphics*[width=8cm]{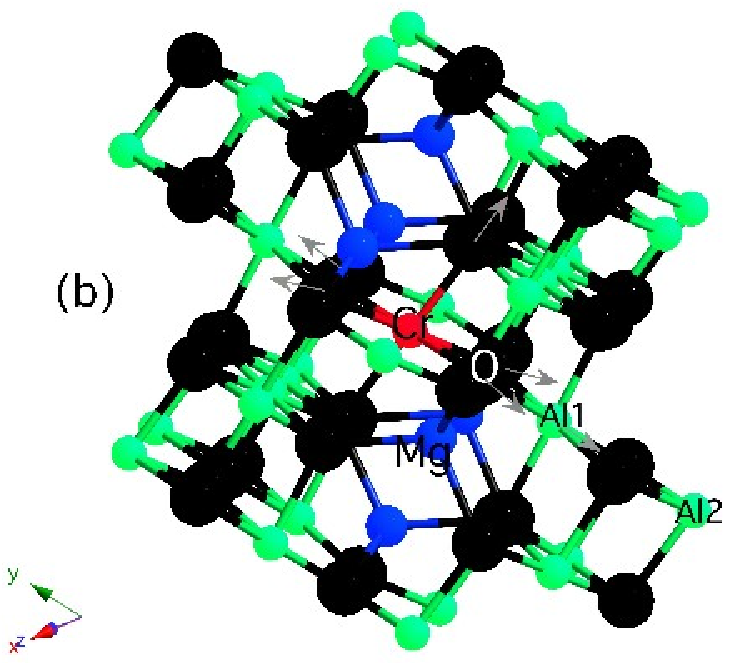}
\caption{\label{fig:3} (color online) (a) Cr-centred octahedron before relaxation (green) and after (red). (b) Model of structural distortions around Cr (red) in MgAl$_{2}$O$_{4}$: Cr$^{3+}$. The O first neighbors (black) and the Al1 (green) second neighbors are displaced outward the Cr dopant in the direction of arrows.  }
\label{fig:site}
\end{figure}

Calculated and experimental interatomic distances are in good agreement (Table~\ref{tab:distances}), a confirmation of the EXAFS-derived radial relaxation around Cr$^{3+}$  after substitution. The symmetry of the relaxed Cr-site is retained from the Al-site in MgAl$_{2}$O$_{4}$ and is similar to the Cr-site in MgCr$_{2}$O$_{4}$. It belongs to the  D$_{3d}$  point group, with an inversion center, three binary axes and a C$_3$ axis (Fig.~\ref{fig:site}a).  This result is consistent with optical absorption\cite{Wood} and Electron-Nuclear Double Resonance experiments\cite{Bravo} performed on MgAl$_{2}$O$_{4}$: Cr$^{3+}$. Our first-principles calculations also agree with a previous investigation of the first shell relaxation, using Hartree-Fock formalism on an isolated cluster.\cite{Votya} 
As it has been mentioned previously, the simulation can provide complementary distances (Fig.~\ref{fig:site}b):  the Al1-O distances, equal to 1.91~\AA, are slightly smaller than Al-O distances in MgAl$_{2}$O$_{4}$. The Al1-Al2 distances are equal to 2.85~\AA, which is close to the Al-Al distances in MgAl$_{2}$O$_{4}$.

Apart from the radial structural modifications around Cr, significant angular deviations are observed in the doped structure. Indeed, the Cr-centred octahedron is slightly more distorted in MgAl$_{2}$O$_{4}$: Cr$^{3+}$, with six O-Cr-O angles of 82.1$^{\circ}$ (and six supplementary angles of 97.9$^{\circ}$): O-Cr-O is more acute than O-Cr-O in MgCr$_{2}$O$_{4}$ (84.5$^{\circ}$, derived from refined structure)\cite{Hill} and than O-Al-O in MgAl$_{2}$O$_{4}$ (either calculated in the present work, 83.5$^{\circ}$, or derived from refined structure, 83.9$^{\circ}$) (Fig.~\ref{fig:site}a).
At a local scale around the dopant, the sequence of edge-sharing octahedra is hardly modified by the substitution (Fig.~\ref{fig:site}b): the Cr-O-Al1 angles (95.1$^{\circ}$) are similar to Cr-O-Cr in MgCr$_{2}$O$_{4}$ (95.2$^{\circ}$) and Al-O-Al in MgAl$_{2}$O$_{4}$ (95.8$^{\circ}$). However, the six Al-centred octahedra connected to the Cr-octahedron are slightly distorted (with six O-Al1-O angles of 86.7$^{\circ}$), compared to O-Cr-O angles in MgCr$_{2}$O$_{4}$ (84.5$^{\circ}$) and O-Al-O angles in MgAl$_{2}$O$_{4}$ (83.9$^{\circ}$).
This modification affects in a similar way the three types of  chains composed of edge-sharing octahedra, in agreement with the conservation of the C$_{3}$ axis. On the contrary, the relative tilt angle between the Mg-centred tetrahedra and the Cr-centred octahedron is very different in MgAl$_{2}$O$_{4}$: Cr$^{3+}$ (with Cr-O-Mg angle of 117.4$^{\circ}$) than in MgCr$_{2}$O$_{4}$  and MgAl$_{2}$O$_{4}$ (with respectively, Cr-O-Mg and Al-O-Mg angles of 124.5$^{\circ}$ and 121.0$^{\circ}$) 
\begin{figure}[!t]
\includegraphics[width=8cm]{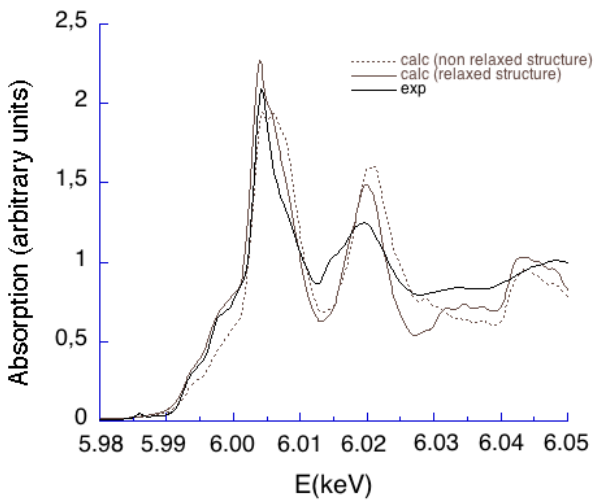}
\caption{\label{fig:5} Cr K-edge XANES spectra in MgAl$_{2}$O$_{4}$: Cr$^{3+}$. The experimental signal (thick line) is compared with the theoretical spectra calculated in the relaxed structure (solid line) and in the non-relaxed structure (dotted line)}
\label{fig:XANES}
\end{figure}

The experimental XANES spectrum of natural MgAl$_{2}$O$_{4}$: Cr$^{3+}$ is shown in Fig.~\ref{fig:XANES}. It is similar to that of a synthetic Cr-bearing spinel. \cite{Levy2} A good agreement with the one calculated from the ab initio relaxed structure is obtained, particularly in the edge region : the position, intensity and shape of the strong absorption peak (peak c) is well reproduced by the calculation. The small features (peaks a and b) exhibited at lower energy are also in good agreement with the experimental ones. In our calculation, the pre-edge features (visible at 5985 eV on the experimental data) cannot be reproduced, since we only considered the electric dipole contribution to the X-ray absorption cross-section:  indeed, as it has been said previously, the Cr-site is centrosymmetric in the relaxed structure, which implies that the pre-edge features are due to pure electric quadrupole transitions. 
The sensitivity of the XANES calculation to the relaxation is evaluated by computing the XANES spectrum for the non-relaxed supercell, in which one Cr atom substitutes an Al atom in its exact position. The result is plotted in Fig.~\ref{fig:XANES}: the edge region (peaks a, b and c) is clearly not as well reproduced as in the relaxed model, and peak e is not visible at all. Therefore, we can conclude that the structural model obtained from our ab initio relaxation is reliable.

The Cr-O distance is larger than the Al-O distance in MgAl$_{2}$O$_{4}$, but is similar to the Cr-O distance in MgCr$_{2}$O$_{4}$ (Table II). This demonstrates the existence of an important structural relaxation around the substitutional Cr$^{3+}$  ion, which is expected since Cr$^{3+}$  has a larger ionic radius than Al$^{3+}$  (0.615~\AA~vs 0.535~\AA).\cite{Shan}  The size mismatch generates indeed a local strain, which locally expands the host structure. As a result, the O atoms relax outward the Cr defect. This radial relaxation is accompanied with a slight angular deviation of the O first neighbors, as compared to the host structure. The magnitude of the radial relaxation may be quantified by a relaxation parameter $\zeta$, defined by the relation: \cite{Mart}
\begin{equation}
\zeta = \frac{\mathrm{R_{Cr-O}}(\mathrm{MgAl_{2}O_{4}:Cr^{3+}})-\mathrm{R_{Al-O}}(\mathrm{MgAl_{2}O_{4}})}{\mathrm{R_{Cr-O}}(\mathrm{MgCr_{2}O_{4}})-\mathrm{R_{Al-O}}(\mathrm{MgAl_{2}O_{4}})}
\end{equation}

We find $\zeta$ = 0.83 (taking the Cr-O experimental distance), close to the full relaxation limit ($\zeta$ = 1), which is more than in ruby $\alpha$-Al$_{2}$O$_{3}$: Cr$^{3+}$  ($\zeta$ = 0.76).\cite{Gaud1}  Vegard's law, which corresponds to $\zeta$ = 0, is thus not obeyed at the atomic scale. The Cr-Al distance is intermediate between the Al-Al and Cr-Cr distances in MgAl$_{2}$O$_{4}$ and MgCr$_{2}$O$_{4}$, which accounts for a partial relaxation of the second neighbors, but the third and fourth shells (O, Mg) do not relax, within the experimental and computational uncertainties. 
The chains of Al-centred octahedra are radially affected only at a local scale around Cr: the Al second neighbors relax partially outward Cr, with a Al1-O bond slightly shortened. The angular deviations are also moderate (below 1$^{\circ}$), since the sequence of octahedra is not modified, but these Al-centred octahedra are slighlty distorted. Indeed, these octahedra being edge-shared, the number of degrees of freedom is reduced, and the polyhedra can either distort or tilt a little, one around another. It is interesting to point out that the three chains of octahedra are orientated along the three four-fold axes of the cubic structure, which are highly symmetric directions.
On the contrary, an angular relaxation (3.5$^{\circ}$) is observed for the Mg atoms, but with the absence of radial modifications. This must be connected to the fact that the tetrahedra share a vertex with the Cr-centred octahedron, a configuration which allows more flexibility for relative rotation of the polyhedra.

The extension of the relaxation process up to the second shell is not observed in the corundum solid solution, in which it is limited to the first coordination shell.\cite{Gaud2} Such a difference between these two solid solutions can be related to the lattice rigidity: the bulk modulus B is smaller in MgAl$_{2}$O$_{4}$ than in $\alpha$-Al$_{2}$O$_{3}$, 200~GPa and 251~GPa, respectively.\cite{Ander} This difference directly arises from the peculiarity of the structure of these two crystals: in the spinel structure, one octahedron is edge-shared to 6 Al octahedra and corner-shared to 6 Mg-centred tetrahedra (Fig.~\ref{fig:site}b). In corundum, each octahedron is face-shared with another, in addition to corner and edge-sharing bonds: this is at the origin of the rigidity of the corundum structure, which is less able to relax around a substitutional impurity such as Cr$^{3+}$, and relaxation is thus limited to the first neighbors.

\section{Conclusions}
This study provides a direct evidence of the structural relaxation during the substitution of Cr for Al in MgAl$_{2}$O$_{4}$ spinel. The local structure determined by X-ray Absorption Spectroscopy and first-principles calculations show similar Cr-O distances and local symmetry in dilute and concentrated spinels. This demonstrates that, at the atomic scale, Vegard's law is not obeyed in the MgAl$_{2}$O$_{4}$-MgCr$_{2}$O$_{4}$ solid solution. Though this result has been obtained in other types of materials (semi-conductors, mixed salts), it is particularly relevant for oxides like spinel and corundum: indeed, the application of Vegard's law has long been a structural tool to interpret, within the so-called "point charge model", \cite{Burns} the color of minerals containing transition metal ions. In spinel, the full relaxation of the first shell is partially accomodated by strain-induced bond buckling, which was found to be weak in corundum: important angular tilts of the Mg-centred tetrahedra around the Cr-centred octahedron have been calculated, while the angles between Cr- and Al-bearing edge-sharing octahedra are hardly affected. The improved thermal and mechanical properties of Cr-doped spinel may be explained by remanent local strain fields induced by the full relaxation of the structure around chromium, as it has been observed in other solid solutions.\cite{Lau} Another important consequence of relaxation concerns the origin of the partition of elements between minerals and liquids in geochemical systems.\cite{Blun} Finally, the data obtained in this study will provide a structural basis for discussing the origin of color in red spinel and its variation at high Cr-contents. Indeed, the origin of the color differences between Cr-containing minerals (ruby, emerald, red spinel, alexandrite) is still actively debated.\cite{Garcia,Garcia2,Gaud1}
\begin{acknowledgments}
The authors are very grateful to O. Proux (FAME beamline) for help during experiment. The theoretical part of this work was supported by the French CNRS computational Institut of Orsay (Institut du D\'eveloppement et de Recherche en Informatique Scientifique) under project 62015. This work has been greatly improved through fruitful discussions with E. Balan, F. Mauri, M. Lazzeri and Ph. Sainctavit. This is IPGP Contribution n$^{\circ}$XXXX. 
\end{acknowledgments}

\end{document}